\RequirePackage{fixltx2e}
\documentclass[aps,prx,reprint,twocolumn,amsmath,
amssymb,amsfonts,longbibliography,superscriptaddress]{revtex4-1}

\usepackage{tikz}
\usepackage{tikz-cd}
\usepackage{amsmath}
\usepackage{hyperref}
\usepackage{color}

\usepackage{microtype}

\usetikzlibrary{positioning}

\usepackage{times}
\usepackage[varg]{txfonts}

\usepackage{mathrsfs}
\usepackage{color}
\usepackage{graphicx}
\usepackage[percent]{overpic}

\usepackage{bm}
\usepackage{fixltx2e}
\usepackage{natbib}

\usepackage{nicefrac}

 \newcommand{\ket}[1]{\left| #1  \right \rangle}

\newcommand{\avg}[1]{\langle #1 \rangle}

\newcommand{\cc}[1]{{#1}^{*}}\newcommand{\cb}[1]{\bar{#1}}

\newcommand{\up}{\uparrow}
\newcommand{\down}{\downarrow}

\newcommand{\figlett}[1]{#1}
\newcommand{\figtitle}[1]{\textbf{#1:}}
\newcommand{\ssection}[1]{\section{#1}}

\newcommand{\kb}{k_{B}}

\newcommand{\tsub}[1]{\textsubscript{#1}}

\newcommand{\abo}[2]{#1\tsub{2}#2\tsub{2}O\tsub{7}}

\newcommand{\rto}[1]{\abo{#1}{Ti}}

\newcommand{\avgtwo}[1]{\avg{\avg{#1}}}
\newcommand{\avgthree}[1]{\avg{\avgtwo{#1}}}

\renewcommand{\vec}[1]{\bm{\mathbf{#1}}}

\newcommand{\vhat}[1]{\vec{\hat{#1}}}

\newcommand{\hhh}{(\nicefrac{1}{2},\nicefrac{1}{2},\nicefrac{1}{2})}

\newcommand{\ham}{H}

\newcommand{\suppinfo}{Supplemental information}

\newcommand{\s}{\sigma}
\newcommand{\eQ}{P}
\newcommand{\exJ}{J}
\newcommand{\exJp}{J_2}
\newcommand{\exJpp}{J_{3a}}
\newcommand{\exJe}{J'}
\newcommand{\SI}{SI}
\newcommand{\ESI}{ESI}
\renewcommand{\SS}{SS}

\newcommand{\onetet}{\vcenter{\hbox{\resizebox{0.2cm}{0.2cm} {
\begin{tikzpicture}[scale=1]
    \coordinate (I1) at (0,0);
    \coordinate (I2) at (1,1);
    \coordinate (I3) at (2,0);
    \coordinate (I4) at (1,-1);
    \draw[black!30!white,line width=1.5mm] (I1) -- (I3);
    \draw[black,line width=1.5mm,rounded corners=0.5mm]
    (I1) -- (I2) -- (I3) -- (I4) -- (I1);
    \draw[black!80!white,line width=1.5mm,rounded corners=0.5mm]
    (I4) -- (I2);
\end{tikzpicture}}}}
}

\newcommand{\twotet}{\vcenter{\hbox{\resizebox{0.4cm}{0.2cm} {
\begin{tikzpicture}[scale=1]
    \coordinate (I1) at (0,2);
    \coordinate (I2) at (0,0);
    \coordinate (I3) at (2,0);
    \coordinate (I4) at (2,2);
    \coordinate (I5) at (1,1);
    \coordinate (I6) at (-1,1);
    \coordinate (I7) at (3,1);
    \draw[black!30!white,line width=1.5mm] (I5) -- (I7);
    \draw[black!30!white,line width=1.5mm] (I1) -- (I2);
    \draw[black,line width=1.5mm,rounded corners=0.5mm]
    (I1) -- (I6) -- (I2) -- (I6) -- (I5) -- (I1) -- (I5) --(I2);
    \draw[black,line width=1.5mm,rounded corners=0.5mm]
    (I3) -- (I7) -- (I4) -- (I7) -- (I3) -- (I4) -- (I5) -- (I3);
\end{tikzpicture}}}}
}

\newcommand{\twoteti}{\vcenter{\hbox{\resizebox{0.42cm}{0.28cm} {
\begin{tikzpicture}[scale=1]
    \coordinate (I1) at (0,2);
    \coordinate (I2) at (0,0);
    \coordinate (I3) at (2,0);
    \coordinate (I4) at (2,2);
    \coordinate (I5) at (1,1);
    \coordinate (I6) at (-1,1);
    \coordinate (I7) at (3,1);
    \draw (I5) node [below=0.5cm] {\fontsize{1.5cm}{1em}\selectfont $i$};    
    \draw[black!30!white,line width=1.5mm] (I5) -- (I7);
    \draw[black!30!white,line width=1.5mm] (I1) -- (I2);
    \draw[black,line width=1.5mm,rounded corners=0.5mm]
    (I1) -- (I6) -- (I2) -- (I6) -- (I5) -- (I1) -- (I5) --(I2);
    \draw[black,line width=1.5mm,rounded corners=0.5mm]
    (I3) -- (I7) -- (I4) -- (I7) -- (I3) -- (I4) -- (I5) -- (I3);
    \draw [fill,black] (I5) circle [radius=0.2];
\end{tikzpicture}}}}
}

\definecolor{cblue}{RGB}{55,126,184}
\hypersetup{colorlinks=true,linkcolor=cblue,citecolor=cblue,urlcolor=cblue} 

\begin{document}
 
\title{Spin slush in an extended spin ice model}
\author{Jeffrey G. Rau}
\affiliation{Department of Physics and Astronomy, University of
Waterloo, Ontario, N2L 3G1, Canada} 
\author{Michel J. P. Gingras} 
\affiliation{Department of Physics and Astronomy,
University of Waterloo, Ontario, N2L 3G1, Canada}
\affiliation{Perimeter Institute for Theoretical Physics, Waterloo,
Ontario, N2L 2Y5, Canada} 
\affiliation{Canadian Institute for Advanced
Research, 180 Dundas Street West, Suite 1400, Toronto, ON, M5G 1Z8,
Canada}
\date{\today}

\begin{abstract}
  We introduce a new classical spin liquid on the pyrochlore lattice
  by extending spin ice with further neighbour interactions. We find
  that this disorder-free spin model exhibits a form of dynamical
  heterogeneity with extremely slow relaxation for some spins while
  others fluctuate quickly down to zero temperature. We thus call this
  state ``spin slush'', in analogy to the heterogeneous mixture of solid
  and liquid water. This behaviour is driven by the structure of the
  ground state manifold which extends the celebrated the
  two-in/two-out ice states to include branching structures built from
  three-in/one-out, three-out/one-in and all-in/all-out tetrahedra
  defects. Distinctive liquid-like patterns in the spin correlations
  serve as a signature of this intermediate range order. Possible
  applications to materials as well the effects of quantum tunneling
  are discussed.
\end{abstract}

\maketitle

The physics of glasses plays an important role in many types of
physical systems; from its origins in the physics of liquids
\cite{berthier-2011-reviews-modern} further realizations have been found
in disordered magnets \cite{binder-1986-reviews-glass},
superconductors \cite{blatter-vinokur-1994-vortices} and metals
\cite{electron-glass-2011-amir-oreg} through to soft-condensed matter
\cite{hunter-weeks-2012-colloidal} and even biophysics
\cite{kirkpatrick-2015-random-first}. While ubiquitous, a complete
understanding of glasses remains an important open problem in
condensed matter physics. Connections between these vastly different
contexts have proven fruitful in making progress; for example, 
studying conceptually and computationally simpler spin models, may
inform the physics of super-cooled liquids and structural
glasses \cite{kirkpatrick-2015-random-first}. However, there are
complications -- while spin glasses are driven by the combination of
random quenched disorder and frustration, glass-forming liquids are
intrinsically \emph{disorder-free}. Finding a disorder-free spin model
that realizes the diverse range of phenomena observed in glass
formers, such as the dramatic slowing down of relaxation and emergence
of spatially heterogeneous dynamics, is a serious challenge. Some examples
of disorder-free spin models with strong freezing have been proposed
\cite{topological-spin-glass-1993-chandra,
lipowski-1997-glassy,jack-berthier-2005-static,chamon-2005-glassiness,
canals-2012-freezing,klich-2014-exotic}.  Each of these proposals has
some deficiency; be it the lack of heterogeneous dynamics, the need
for multi-spin interactions, the use of uncontrolled approximations or
the introduction of non-local dynamics.

\begin{figure}[t]
  \centering
  \includegraphics[width=0.8\columnwidth]{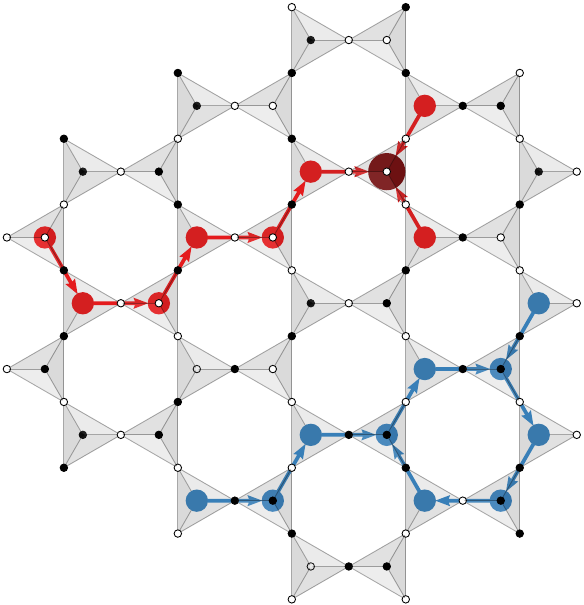}
  \caption{
    \label{fig:kagome-plane}
    \figtitle{Example of spin slush ground state} A
    spin slush ground state that includes all instances of the rules discussed
    in the main text. 
    The colours indicate $\sigma_i = \pm 1$ (black, 
    white) for the pyrochlore sites, and the charge $Q_I$ for the dual
    lattice with $Q_I=0$ (gray), $Q_I = \pm 1$ (red, blue) and $Q_I = \pm
    2$ (dark red, dark blue).  The arrow passes through the location of the
    minority spin for a single charge.  This state contains branching
    lines of charge of both signs, a charge loop and a double charge tetrahedron.
  }
\end{figure}

In this article, we introduce a new type of cooperative paramagnet
which we call \emph{spin slush} (\SS{}) in an extended spin ice model (\ESI{}). 
This classical \SS{} model is disorder-free
and includes only first-, second- and third-neighbour Ising bilinear
exchange interactions and thus lacks the pitfalls
discussed above. We start from spin ice (\SI{}), 
a well-studied magnetic analogue of common water ice \cite{bramwell-2001-science},
magnetic moments pointing in or out of
the corner-shared tetrahedra of the pyrochlore lattice embody the
proton displacements of water ice
\cite{harris-zinkin-1996-frustration}. Similar to water ice, spin ice
displays an extensive ground state degeneracy, and thus an associated
extensive residual entropy, characterized by the two-in/two-out ``ice
rule'' condition on each tetrahedron \cite{bramwell-2001-science}. 
In SS, we find that the
ground state manifold of \SS{} is \emph{larger} than that of \SI{} and contains a
far richer set of states. In addition to the two-in/two-out tetrahedra
of the spin ice ground state manifold, there are spatially extended
structures assembled from three-in/one-out, three-out/one-in and
all-in/all-out tetrahedra.  Built from \SI{} defects, these structures are not
simply loops or strings, but include branching tree-like features.
After characterizing the static thermodynamic and magnetic properties
of \SS{}, we turn to dynamics. Approaching zero temperature, we find
freezing, as in \SI{}
\cite{snyder-cava-2001-freezes,jaubert-holdsworth-2009-signature},
with an exponentially increasing average relaxation time. However,
unlike in \SI{} where all of the spins freeze uniformly as the
temperature is lowered, the spins in the \SS{} exhibit highly
heterogeneous dynamics reminiscent of glass formers
\cite{berthier-2011-dynamical}.  While many of the spins strongly
freeze with an extremely slow relaxation rate, a fraction of the
spins, organized into spatially local clusters, remain completely
dynamic, relaxing almost immediately. Since this model is
disorder-free, the random distribution of these dynamical spins
derives solely from the overall freezing behaviour.  This dynamical
heterogeneity in \SS{} at low temperatures motivates the name ``spin
slush'', in analogy to ``slush'' where liquid water and solid ice
coexist as a mixture.  Finally, we speculate on the behaviour of
``quantum spin slush'' as well as possible experimental relevance in
frustrated pyrochlore magnets.

\ssection{Model}
We start with a review of some results for the nearest-neighbour \SI{}
model \cite{ordering-1956-anderson} to establish our notation and motivate 
the \SS{} model. The \SI{} model is a nearest-neighbour Ising antiferromagnet on the
pyrochlore lattice, $J \sum_{\avg{ij}} \s_i\s_j$, where $\s_i = \pm 1$ are the
Ising spins. This can be reformulated in terms of ice rule defects, or \emph{charges},
defined on each tetrahedron. With each tetrahedron identified with a
dual diamond lattice site $I$, one defines the charge $Q_I \equiv
\frac{1}{2}(-1)^I \sum_{i \in I} \s_i$, where $(-1)^I$ is a sign
reflecting the sublattice of $I$.  In this language, the
nearest-neighbour \SI{} Hamiltonian simply penalizes non-zero charges,
taking the form
\begin{subequations}
  \label{eq:si}
\begin{align}
  \ham_{\rm SI}
  &\equiv \exJ \sum_{\avg{ij}} \s_i\s_j
  = 8\exJ
    \sum_{\onetet{}} \left(\frac{1}{2} \sum_{i \in \onetet{}} \s_i \right)^2 -N \exJ, \\
  &= 2\exJ \sum_I Q_I^2 - N \exJ.
\end{align}
\end{subequations}
The ground states of this model are those with $Q_I=0$ for all
tetrahedra, i.e. the celebrated two-in/two-out states of the ice
manifold. This manifold is macroscopically degenerate with a residual
entropy given approximately by $S_{\rm SI} \sim (N \kb/2) \log{(3/2)} \sim
0.202 N\kb$ \cite{bramwell-2001-science}.  Due to this extensive ground
state degeneracy, addition of small perturbations will generically
select an ordered state from this manifold at low temperatures
\cite{bramwell-2001-science}.
\begin{figure}[tp]
  \centering
  \includegraphics[width=\columnwidth]{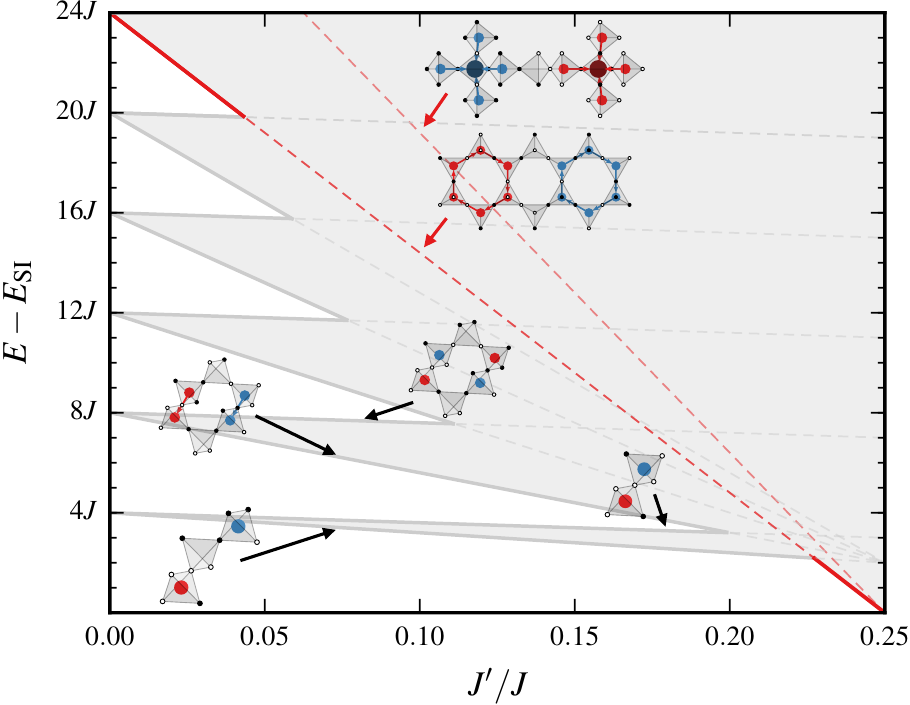}
  \caption{\label{fig:excitations}
    \figtitle{Collapse of excitations in extended spin ice}    
    We sketch the structure of the excited states of the
    model of Eq. (\ref{eq:model}) along the $J_2 = J_{3a} \equiv J'$ line.
    When $J'$ is finite, the highly degenerate bands of single and double charge
    states are split due to the nearest-neighbour attraction embodied in the
    second term in Eq. (\ref{eq:esi-alt}).
    For the low lying bands, we illustrate the charge arrangements that 
    are favoured and those that are disfavoured by $J'$, using the notation
    of Fig. \ref{fig:kagome-plane}. Near the spin slush at 
    $J'/J \sim 1/4$ an infinite set of excited states collapse to zero energy. 
    We have illustrated two of the simplest examples, built from twelve charges,
    with energy shown by red lines.
  }
\end{figure}

To explore the effects of such perturbations, we consider the addition
of second and third-neighbour Ising exchanges of the form
\begin{equation}
\label{eq:model}
  \ham \equiv 
\exJ \sum_{\avg{ij}} \s_i \s_j+
\exJp \sum_{\avgtwo{ij}} \s_i \s_j+
\exJpp \sum_{\avgthree{ij}_a} \s_i \s_j.
\end{equation}
We include only the third-neighbour exchanges that are composed of two
nearest-neighbour steps. For many mechanisms that generate such
interactions, for example super-exchange or through virtual crystal
field excitations, one expects the interactions $\exJp$ and $\exJpp$
to be generated on equal footing. The other third-neighbour exchange,
$J_{3b}$, spanning the hexagons of the pyrochlore lattice, is only
generated at higher order. Significant second- and third-neighbour
exchange can be present in real materials either intrinsically 
\cite{cheng-2008-long-range-density-functional,yaresko-2008-exchange},
or via partially cancellation of the leading terms \cite{molavian-gingras-2007-dynamically}.
One can show that for any spin ice state
\begin{equation}
\sum_{\avgtwo{ij}} \s_i \s_j + \sum_{\avgthree{ij}_a} \s_i \s_j = {\rm const.}
\end{equation}
We thus see that two terms are not independent
and when $\exJp=\exJpp \equiv \exJe $ they
\emph{cancel} each other when in the SI manifold.  Moving along the $\exJp =
\exJpp$ line, the model moves away from the nearest-neighbour \SI{} regime, but without
lifting the degeneracy of the ice manifold.  While \SI{} persists
as the ground state at low temperature for sufficiently small $\exJe/\exJ$,
eventually it gives way when another set of states crosses the
\SI{} manifold \footnote{ This line of degeneracy also exists for
analogous models with $N$-components spins, though how the termination
manifests depends on the precise value of $N$.}.  We thus refer to the model along this line
as \emph{extended spin ice} (ESI).  It will prove useful to write
this model in terms of the charges $Q_I$ as
\begin{equation}
\label{eq:esi-alt}
\ham_{\rm \ESI{}} =
2(\exJ-2\exJe) \sum_I Q_I^2  - 4 \exJe \sum_{\avg{IJ}} Q_IQ_{J} - N(\exJ - \exJe).
\end{equation}
We see that $\exJe >0$ generates an attraction between
nearest-neighbour charges of the \emph{same} sign.  This short-range
attraction between charges will play a central role in understanding
the ground and excited states of \ESI{}.

One can show \cite{supp} that the \SI{} manifold persists until
$\exJe=\exJ/4$ for $\exJe > 0$ and to $\exJe=-\exJ/2$ for
$\exJe < 0$ \footnote{
At $\exJe = -\exJ/2$, the ground states include all configurations
with staggered charge $Q_I = Q_0(-1)^I$. These are the ice states
($Q_0=0)$, the single charge states ($Q_0 = 1$) and the all-in,
all-out states ($Q_0 = 2$).  The manifold of $Q_I = (-1)^I$ states has
been discussed recently in Ref.~[\onlinecite{jaubert-2015-holes}],
albeit starting from a very different model.
}
The collapse of excited states when approaching
$J' = J/4$ is illustrated in Fig. \ref{fig:excitations}. We show only
the simplest examples that cross the ice manifold, but as we shall
see, there are an \emph{infinite} set of such states.  We focus on the
end-point at $\exJe = \exJ/4$ which we will refer to as the \SS{}
model. At this special point one can write the model as
\begin{subequations}
  \label{eq:esi}
  \begin{align}
    \ham_{\rm \SS{}}
    &\equiv \exJ \sum_{\avg{ij}} \s_i \s_j+
      \frac{\exJ}{4} \sum_{\avgtwo{ij}} \s_i \s_j+
      \frac{\exJ}{4} \sum_{\avgthree{ij}_a} \s_i \s_j, \\
    &= \frac{\exJ}{2} \sum_{\twotet{}} \left(\frac{1}{2} \sum_{i \in\twotet{}} \s_i \right)^2 
      -\frac{7N \exJ }{8}.
\end{align}
\end{subequations}
In this form, one notes a strong similarity to the \SI{} model of
Eq. (\ref{eq:si}), except with the fundamental unit being a
\emph{pair} of tetrahedra, indicated by $\twotet{}$, rather than a single tetrahedron.

\ssection{Ground state manifold} 
The ground state manifold of \SS{} is most easily characterized in
terms of the variables
\begin{equation}
  \eQ_{i} \equiv \frac{1}{2} \sum_{\overset{j \in}{} \twoteti{}} \s_j.
\end{equation}
Following Eq. (\ref{eq:esi}), any state with $\eQ_i =
\pm\nicefrac{1}{2}$ for all sites has the minimal
energy $-3N\exJ/4$ and is in the ground state manifold.
Alternatively, we can write this in terms of the \SI{} charges;
associating each site $i$ of the pyrochlore lattice with a
nearest-neighbour bond $\avg{IJ}$ of the dual diamond lattice, one has
$\eQ_{i} = (-1)^I(Q_I - Q_{J}) - \s_i/2$.  From this expression for $\eQ_i$ in
terms of the \SI{} charges, it is clear
that any \SI{} state with $Q_I=0$ for all sites also belongs to the
\SS{} manifold.  In addition to the familiar \SI{} states, many more
states satisfy $\eQ_i = \pm \nicefrac{1}{2}$. A na\"ive enumeration of
states for an isolated pair of tetrahedra shows that beyond the 18 ice
states, there are an additional 52 states, 70 total, that belong to
the \SS{} manifold \footnote{
We caution that a Pauling-like estimate severely underestimates the
degeneracy of the \SS{} manifold. Given that the number of tetrahedron
pairs is equal to the number of sites one would estimate a residual
entropy of  $N \kb (\log{2} + \log{(70/2^7)}) \sim 0.0896 N\kb$. 
This reflects that the constraints provided by $P_i = \pm \nicefrac{1}{2}$
are much less independent than in spin ice where Pauling's estimate is
accurate.
}. These additional states include configurations with both single charge ($Q_I
= \pm 1$) and double charge ($Q_I = \pm 2$) defects.  The influence of
the nearest-neighbour attraction of charges manifests here; pairs of
like single charges can appear together, while double charges only appear
with accompanying single charges of the same sign. One finds
from Eq. (\ref{eq:esi-alt}) that the energy cost of having a charge can be
compensated by the energy gain of having two neighbouring charges of
the same sign.

From these observations, we formulate rules for constructing states
that satisfy $\eQ_i = \pm \nicefrac{1}{2}$. We formulate these rules
from the perspective of specifying non-ice tetrahedra ($Q_I \neq 0$)
states first, then populating the remaining tetrahedra with any
compatible ice states afterward.  The rules for placing the non-ice,
charged tetrahedra are:

\begin{enumerate}
  \item \emph{Single charge rule:}
    The minority spin of a single charge, $Q_I = \pm 1$, must be
    connected to a tetrahedron carrying a single or double charge of the
    same sign.
  \item \emph{Double charge rule:}
    A double charge $Q_I = \pm 2$ must have its four nearest-neighbour 
    tetrahedra occupied by single charges of the same sign.
  \item \emph{Neighbour rule:} 
    A single charge, $Q_I \pm 1$, cannot have any single charges of
    opposite sign as nearest neighbours.
\end{enumerate}
Once single and double charges have been placed such that they satisfy
the above three rules, one can fill the remaining tetrahedra with
\emph{any} allowed ice rule, $Q_I = 0$, states.  The first rule allows
the single charge tetrahedra to form branching tree-like structures
\footnote{ In the language of graph theory, the single charge structures can form
directed graphs without sinks (i.e. each vertex has non-zero out
degree).}, where the minority spin of a given charge also belongs to
the next charge in the structure. Each branch must terminate in some
way compatible with the rules, so the minority spin must end up on
another single charge. The possibilities for terminating a branch include
looping back to itself, ending on a different branch or on one of the
single charges attached to a double charge. Note that these single and
double charge structures must exist for both signs of the charge to
satisfy the global neutrality requirement $\sum_I Q_I = 0$. The third rule
implies that charge structures of opposite sign must be separated by
at least one ice rule obeying tetrahedron.  An illustration of an \SS{} state
incorporating all of these features, restricted to a single $[111]$ 
kagom\'e plane, is shown in Fig. \ref{fig:kagome-plane}.

\begin{figure}[t]
  \centering
  \begin{overpic}[width=\columnwidth]
    {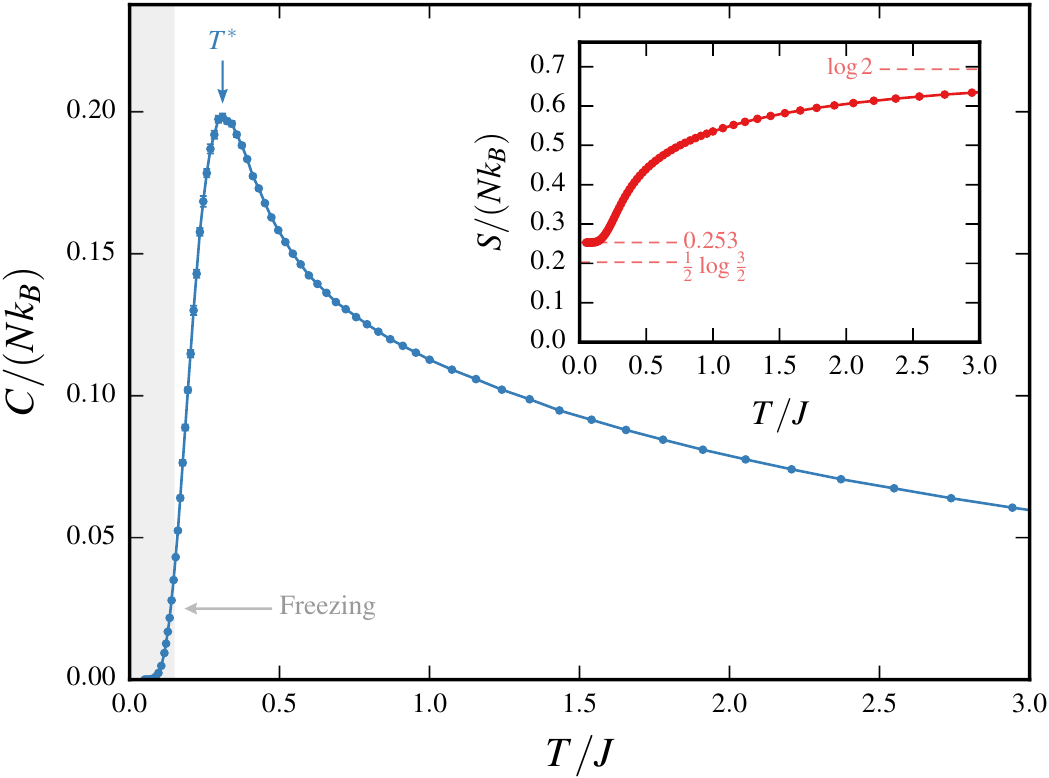}
    \put(93,11.5){\figlett{(a)}}
    \put(87.5,43.5){\figlett{(b)}}
  \end{overpic}
  \caption{\label{fig:thermodynamics}    
    \figtitle{Specific heat and entropy of extended spin ice}
    Finite temperature (a) specific heat, $C$, and (b) 
    entropy, $S$,
    of the spin slush model for a system of $10^3$ conventional cubic cells
    of the pyrochlore lattice.
    Entrance into the spin slush manifold is signaled by the peak 
    in the specific heat at $\cc{T} \sim 0.3 \exJ$. 
    Residual entropy as $T \rightarrow 0$ is
    $S \sim 0.253 N\kb$. Freezing becomes apparent below
    $T \sim 0.15 \exJ$, as indicated by the shaded region.
  }
\end{figure}

\begin{figure}[t]
  \centering
  \begin{overpic}[width=\columnwidth]
    {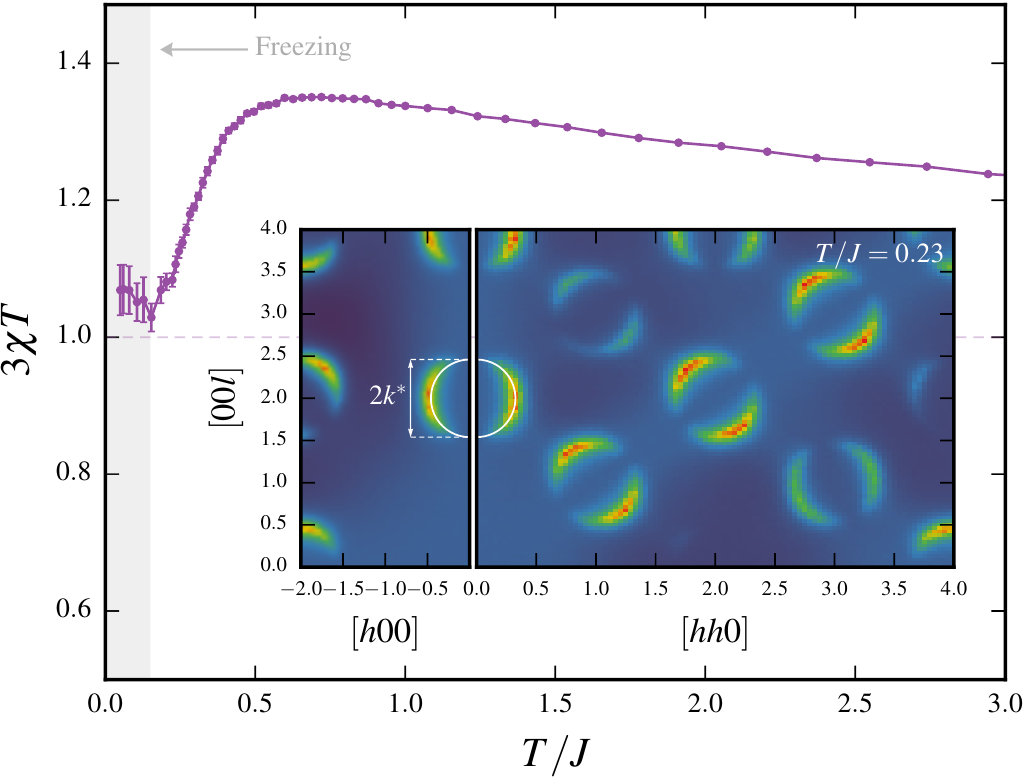}
    \put(92,71){\textcolor{black}{\figlett{(a)}}}
    \put(88,22){\textcolor{white}{\figlett{(b)}}}
  \end{overpic}
  \caption{\label{fig:suscept}
    \figtitle{Magnetic properties of extended spin ice}
    (a) Finite temperature susceptibility $\chi$ 
    of the spin slush model for a system of $10^3$ cubic cells. The
    susceptibility passes through maximum near $T\sim 0.6 \exJ$
    before settling into a Curie-like regime with $3\chi T \sim 1$.
    Freezing becomes apparent below $T \sim 0.15 \exJ$, with the
    susceptibilities depending on the detailed spin configuration of the frozen state.
    (b) Transverse moment-moment correlation function $I(\vec{k})$ defined in Eq. (\ref{eq:sf}),
    for the spin slush model at $T=0.23J$ for a system of $24^3$ cubic cells.
    Cuts in the $[hhl]$ and $[h0l]$
    planes are shown. Correlations are peaked on
    spherical surfaces of radius $\cc{k} \sim 0.5 (2\pi/a)$ where $a$
    is the size of a cubic unit cell. These spheres are centered on the
    locations of the pinch-points in spin ice.
  }
\end{figure}
 
\ssection{Thermodynamic and magnetic properties}
With the ground states of \SS{} identified, we now explore the finite
temperature properties via classical Monte Carlo simulations using
single-spin flip dynamics augmented with parallel tempering when
appropriate.  Basic
thermodynamic quantities are shown in Fig. \ref{fig:thermodynamics}.
The specific heat exhibits a broad peak at $\cc{T} \sim 0.3 \exJ$,
reminiscent of the peak seen in \SI{}.  This peak signals the
release of entropy as one begins to enter the \SS{} ground state manifold. This
can be seen explicitly in the entropy in Fig. \ref{fig:thermodynamics}
where, below $\cc{T}$, the entropy approaches the constant value $S_{\rm
\SS{}} \sim 0.253 N\kb$. As expected from the rules derived in the
previous section, this is significantly \emph{higher} than $S_{\rm SI} \sim 0.202N\kb$
found in \SI{}.  At these low temperatures severe freezing is encountered,
preventing the simulations from reaching equilibrium below $T\sim 0.15 \exJ$ 
\footnote{
  It is not obvious how to construct a non-local move that samples
  the \SS{} manifold efficiently. Including the \SI{} loop move 
  does aid equilibriation, but it is only effective in regions where no
  single and double charge defects are present, leaving the freezing
  problem for future work.
}.
The frozen states
belong to the \SS{} manifold and exhibit the single and double
charge structures discussed in the previous section. We found
no evidence of ordering in any of our simulations, be it conventional or
via an order-by-disorder mechanism.  Further, the specific heat and
entropy are somewhat immune to this freezing problem, showing
consistent behaviour between simulations. The magnetic properties
however are more sensitive.

The simplest probe of the magnetic behaviour is the uniform
susceptibility, $\chi$, shown in Fig. \ref{fig:suscept}, for the
moments $\vec{\mu}_i \equiv \s_i \vhat{z}_i$ pointing in/out of the
tetrahedra along the local $[111]$ direction $\vhat{z}_i$.  At both
low and high temperatures, one finds Curie-like behaviour, with $3
\chi T$ constant, separated by a broad peak at $T \sim O(J)$.
The constant approached as $T \rightarrow 0$ depends on the details of
how the system freezes.  This varies between simulations, taking on a
distribution of values clustered around $3 \chi T \sim 1$, reflected
in the large error bars in Fig. \ref{fig:suscept}.  A more detailed probe
of the magnetic structure can be obtained from the spin-spin
correlations, as can be directly observed via neutron scattering. 
Recall that in \SI{} the appearance of sharp
``pinch-points'' \cite{isakov-gregor-2004-dipolar} in the transverse
moment-moment correlation function
\begin{equation}
  \label{eq:sf}
  I(\vec{k}) \equiv \frac{1}{N} \sum_{ij} e^{i \vec{k} \cdot (\vec{r}_i-\vec{r}_j)} 
\left[\vhat{z}_i \cdot \vhat{z}_j - 
\left(\vhat{z}_i \cdot \vhat{k}\right)
\left(\vhat{z}_j \cdot \vhat{k}\right)\right]
  \avg{\s_i\s_j},
\end{equation}
signals the development of long-range dipolar spin correlations.  In
\SS{}, one finds sharp features in the spin-spin correlation function
distinct from such pinch points. As shown in Fig. \ref{fig:suscept},
below $\cc{T}$ the spin correlations develop into sharp rings centered
on zone centers in a given plane of reciprocal space. In the full
$[hkl]$ space, these features
lie approximately on \emph{spheres}, reminiscent of an
isotropic liquid.  This analogy is even more striking in the structure
factor of the spin ice charges $Q_I$ where the intensity is approximately uniform
across the sphere \cite{supp}. The wave-vector
$|\vec{k}| \sim 0.5 (2\pi/a) \equiv \cc{k}$, where $a$ is the
size of the conventional cubic unit cell, indicates these correlations
have a characteristic length of $2$ cubic cells and thus represent
intermediate scale correlations. These correlations are consistent
with the typical size of the charged structures that appear in the \ESI{}
manifold. Indeed, as seen in Fig. \ref{fig:kagome-plane}, even the
smaller of these structures can span several cubic unit cells.

\begin{figure}[t]
  \centering
  \begin{overpic}[width=\columnwidth]
    {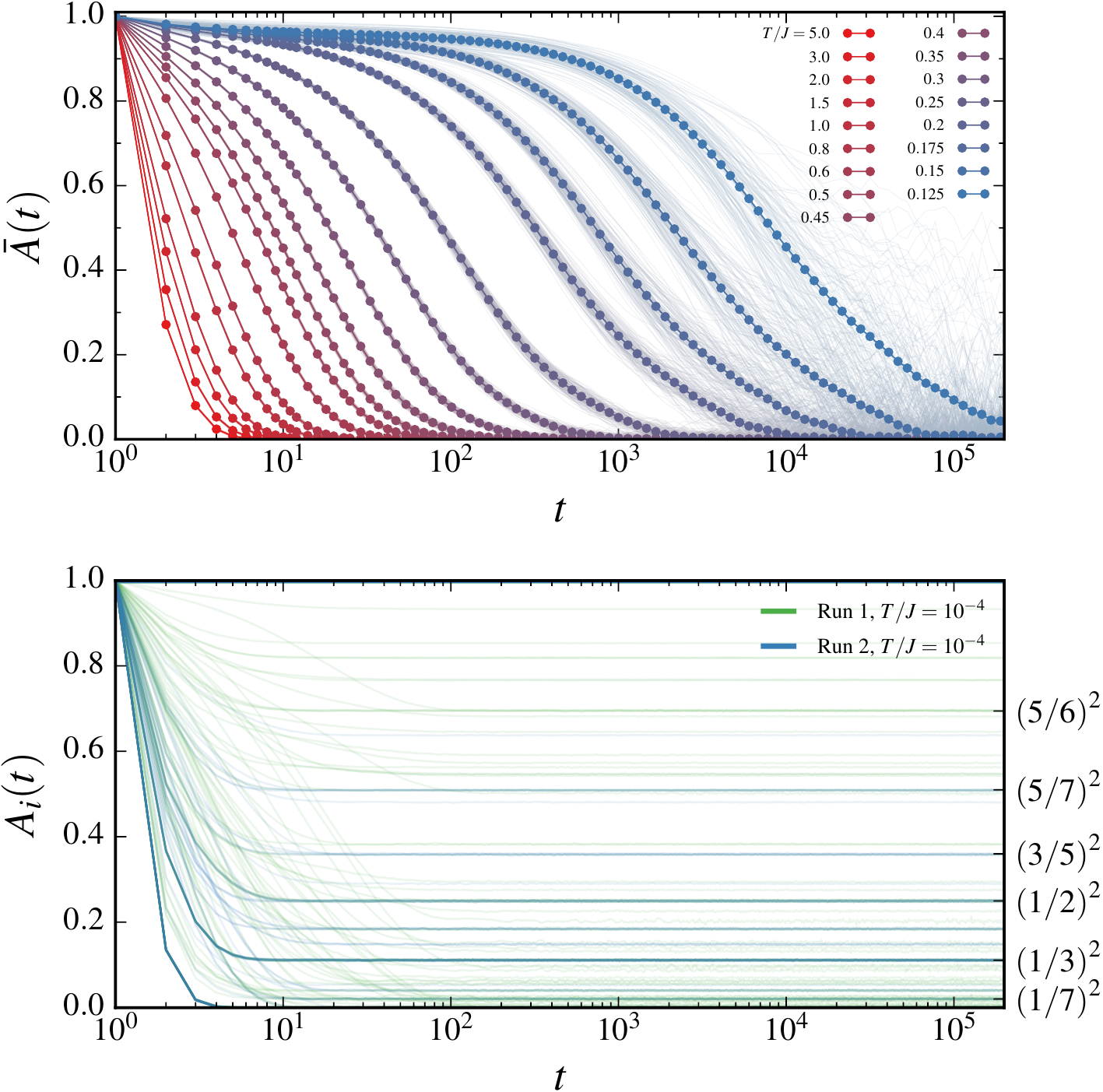}
    \put(10,60){\figlett{(a)}}
    \put(10,8.5){\figlett{(b)}}
  \end{overpic}
  \caption{\label{fig:autocorrelations}    
    \figtitle{Auto-correlation functions in extended spin ice} 
(a) Site-averaged auto-correlation function $\cb{A}(t)$ at various
temperatures for a system of $8^3$ cubic cells. As we approach low
temperatures the relaxation time grows exponentially. Short-time
dynamics is apparent in the initial decrease of $\cb{A}(t)$ for $t
\lesssim 10^2$. The thin curves show a sample of the individual
site-resolved $A_i(t)$ at each temperature, showing increasing levels
of heterogeneity for $T \lesssim \cc{T}$.  (b) Site-resolved
auto-correlation functions $A_i(t)$ at the very low temperature
$T=10^{-4}J$. We show two distinct annealed runs of a system of $8^3$
cubic cells. Aside from essentially frozen spins with $A_i(t) =1$, one
finds many spins that relax over time scales of $10^1$ or $10^2$
sweeps. At long times the auto-correlation functions reach constant
values $A_i(\infty)$ that are clustered about the squares of 
simple, rational numbers (see text).
  }
\end{figure}

\begin{figure*}[t]
  \centering
  \begin{overpic}[width=0.9\textwidth]
    {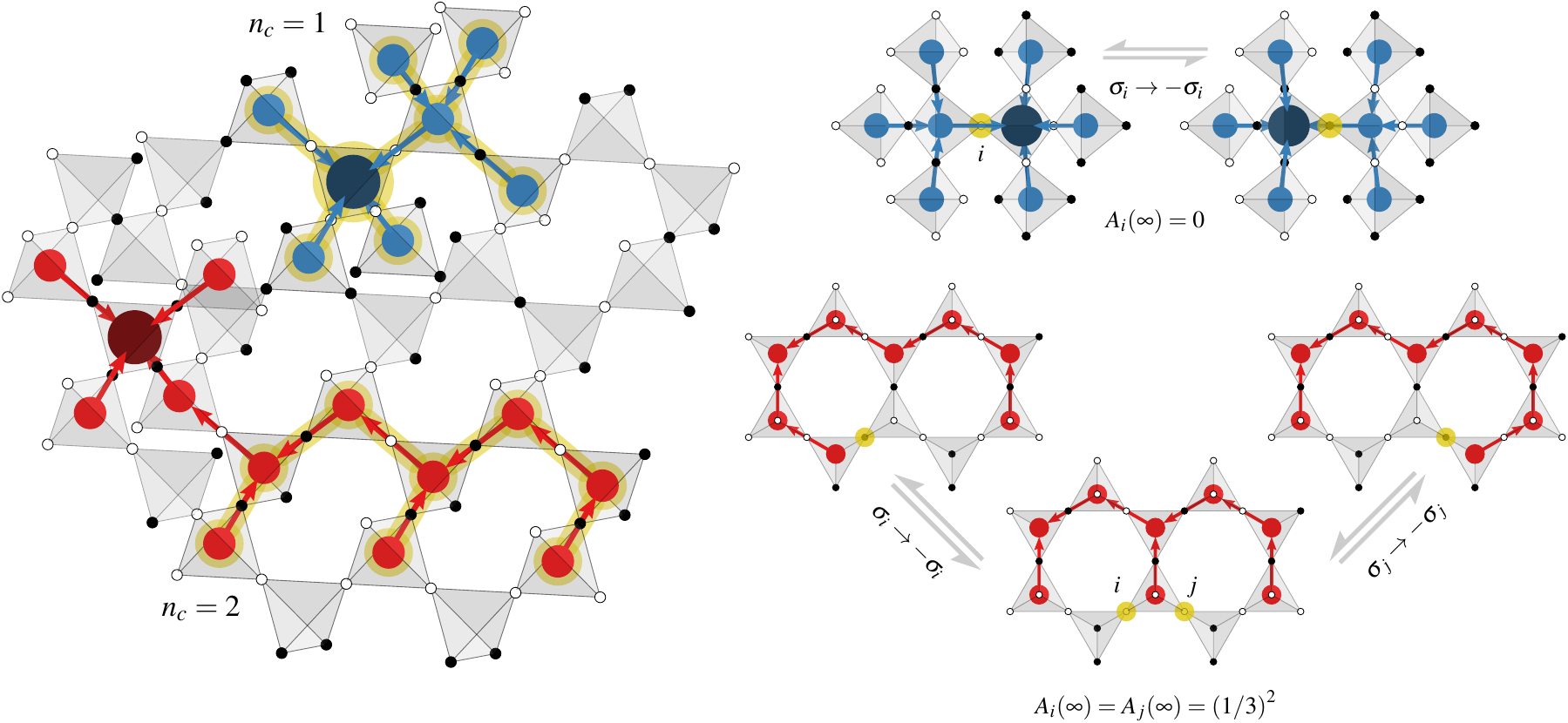}    
    \put(5,38){\figlett{(a)}}
    \put(53,42){\figlett{(b)}}
    \put(49,15){\figlett{(c)}}
  \end{overpic}
  \caption{\label{fig:dynamics}
    \figtitle{Dynamical clusters in spin slush}
We illustrate some of the dynamical clusters that can appear in
the spin slush ground state manifold. In (a) we show an example of part of
a state with two such clusters, one containing a single dynamical spin
($n_c=1$) and the other containing two dynamical spins ($n_c=2$)
highlighted in gold. In (b-c) we show the accessible states of each
dynamical cluster. For $n_c=1$ the there are two states yielding an
average spin of zero and thus $A_i(\infty)=0$.  For the $n_c=2$ case,
one finds three accessible states with an average spin of $\pm 1/3$
and thus $A_i(\infty) = A_j(\infty) = (1/3)^2$.
}
\end{figure*}

These simulations confirm that the \SS{} model does not order and the
\SS{} manifold shows all the rich charge structures at intermediate
length scales implied by the \SS{} rules. Indeed, at low temperatures
a significant fraction of tetrahedra, approximately $30-35\%$,
carry single-charges while a smaller but finite fraction, a percent or so,
carry double charges. Similar to the susceptibility, the amount
of single and double charges present at low temperatures varies
somewhat from run to run, a consequence of the severe freezing problem.
To better understand this issue, we now look
more closely at the low temperature dynamics of \SS{}.

\ssection{Dynamics and ``spin slush''}
To reflect the physics of real
systems with local dynamics, we employ only single spin flip, Metropolis
dynamics, though we expect \emph{any} local dynamics to give
qualitatively the same behaviour.  We primarily consider the
site-resolved auto-correlation functions, defining
$A_i(t) \equiv \avg{\s_i(t_0) \s_i(t_0+t)}$, where $\s_i(t)$ is the
Ising spin at a given Monte Carlo sweep $t$ at site $i$, averaging over many
initial times $t_0$. Generically, one would expect exponential relaxation
$A_i(t) \sim e^{-t/\tau_i}$ with a characteristic relaxation time
$\tau_i$. Indeed this is found in \SI{}, with the relaxation time
being site-independent, with $\tau \sim \tau_i$ and increasing
exponentially as temperature is lowered
\cite{jaubert-holdsworth-2009-signature}.

In contrast to \SI{}, the dynamics in \SS{} vary strongly from site to
site. As temperature is lowered, most of the sites freeze, with their
relaxation times becoming very long, similar to what is found in
\SI{} \cite{snyder-cava-2001-freezes,jaubert-holdsworth-2009-signature}.  
This can be seen in the site-averaged auto-correlation
function $\cb{A}(t)$ shown in Fig. \ref{fig:autocorrelations}.
However, there are clear differences, namely in the initial decrease
and plateau in $\cb{A}(t)$ at short times as well as in the larger
site to site variance in $A_i(t)$ at low temperatures. We can
understand this behaviour by looking at the $T \rightarrow 0$ limit;
one finds that a fraction of sites remain highly dynamic down to very
low temperatures.  This is illustrated in
Fig. \ref{fig:autocorrelations}, where the site-resolved
auto-correlation functions are shown for $T = 10^{-4} \exJ$. The
frozen spins have $A_i(t) = 1$ at all times, while the unfrozen
spins have $A_i(t)$ relaxing in $10^1$ to $10^2$ sweeps to a constant
value $\lim_{t\rightarrow \infty} A_i(t) \equiv A_i(\infty) < 1$
\footnote{ To be precise, for the long-time limit $\lim_{t\rightarrow
\infty} A_i(t)$ we mean $t \gg 1$ but still much smaller than the slow
timescale $\sim O(e^{J/T})$.  }.  A non-zero value of $A_i(\infty)<1$
indicates that, while fluctuating, on average more time is spent in
one of the states $\s_i = \pm 1$ than the other. For example, if
$\sigma_i$ is sampling uniformly from values $\sigma^{(1)}, \dots,
\sigma^{(m)}$ as a function of time, then $A_i(\infty) \sim
(\frac{1}{m}\sum_{n=1}^m \sigma^{(n)})^2$ at long times. 
Fig. \ref{fig:autocorrelations} shows that the long-time values
$A_i(\infty)$ cluster about the squares of rational numbers, as would
be expected from the above discussion. In these annealed simulations, the
frozen spins make up the bulk of the system, while the number of
unfrozen, dynamic spins is on the order of a few percent.

To better understand these dynamic spins, we examine their real space
structure. We find that these spins are spatially correlated, forming
clusters \footnote{A dynamical cluster is defined by a set of spins
  where $\lim_{t \rightarrow \infty} A_i(t) < 1$ and each spin is
  connected by a first, second or third neighbour bond to another spin
  in the cluster.} of varying size $n_c$. The \SS{} state at low
temperature is thus a mixture where regions of frozen and unfrozen
spins coexist. Dynamical clusters built from a small number of sites
can be identified directly from the \SS{}
rules. Fig. \ref{fig:dynamics} shows an \SS{} ground state containing
several of these dynamical clusters. For example, one has a single
site that can be flipped while preserving all of the \SS{} rules,
representing an $n_c=1$ dynamical cluster. A larger $n_c=2$ cluster is
also shown, where two spins can be flipped, though not independently.
For both these examples we note that a large number of surrounding
frozen spins are needed to construct these dynamical clusters. A
na\"ive counting for the $n_c=1$ case yields a fraction of unfrozen to
frozen spins of $\sim 1/25 \sim 4 \%$, comparable to the few percent
average of unfrozen spins observed in our annealed simulations.  These
examples represent only a small subset of the possible dynamical
clusters that can be constructed in the \SS{} manifold.  In the
\suppinfo{} \cite{supp}, we show several ways to construct dynamical
clusters of \emph{arbitrary} size as well as direct illustrations of
the time evolution of dynamical clusters in simulations of small
systems \cite{supp}.  The presence of such dynamical clusters is not
specific to the single-spin-flip dynamics used; for example, analogous
dynamical clusters can be constructed for spin-exchange dynamics
\cite{supp} and we expect the same holds true for \emph{any} local
dynamics.

\ssection{Discussion}
Outside of any pure theoretical interest, one may be concerned with
the fine-tuning required to reach the \SS{} phase. As in \SI{}
\cite{bramwell-2001-science}, though the precise point in phase space
may be difficult to reach in a material realization, the nearby
regions in phase space may be controlled primarily by the \SS{}
physics.  Understanding the \SS{} manifold then allows one to
understand the surrounding phases and their higher temperature
properties as perturbations to the \SS{} model.  Here we discuss two
types of such perturbations: deviations from the $\exJp = \exJpp$
\ESI{} line and quantum terms, such as transverse field or exchange.

While the effects of finite second- and third-neighbour exchange on
similar models has been studied extensively \cite{
  reimers-berlinsky-1991-pyrochlores,
  nakamura-hirashima-2007-classical-pyrochlore,
  chern-moessner-2008-partial, conlon-chalker-2010-pinch,
  ishizuka-2014-monte-carlo-ising}, the regime along the \ESI{} line
and near the \SS{} point remains largely unexplored. We find four
neighbouring phases; the simplest are a \hhh{} ordered phase expected
from the $\exJpp \rightarrow +\infty$ limit that appears for $\exJpp >
\exJ/4$ and a ferromagnetic \SI{} state expected from the $\exJpp
\rightarrow -\infty$ limit that appears for $\exJpp < \exJ/4$.  For
$\exJp < \exJ/4$ one finds a set of layered states with sub-extensive
degeneracy $\sim 2^L$
\footnote{These are related to, but not identical to the layered
  states discussed for the $\exJ$-$\exJp$ classical Heisenberg model
  of Ref. \cite{chern-moessner-2008-partial}}. For $\exJp > \exJ/4$
one finds a complex incommensurate ordering with wave-vector along
$[h00]$ or equivalents.  The \SS{} manifold ties these phases
together, \emph{all} of which are drawn from the \SS{} ground state
manifold, with $\eQ_i = \pm \nicefrac{1}{2}$ for all pairs of
tetrahedra, and extend over large regions of parameter space.  We
leave the detailed investigation of these neighbouring phases and
other perturbations (such as $J_{3b}$, dipolar interactions, etc) for
future studies.

The effect of quantum non-Ising interactions on \SS{} is potentially
much richer than in \SI{}.  In the latter, the addition of transverse
exchange or transverse field induces tunneling within the \SI{}
manifold yielding a $U(1)$ quantum spin liquid (QSL)
\cite{hermele-balents-2004-photons,
  banerjee-isakov-2008-unusual,shannon-2012-quantum-ice,mcclarty-gingras-2014-spin-ice}.
This QSL is described by an emergent electrodynamics, complete with a
gapless photon excitation \cite{hermele-balents-2004-photons}.
However, the associated energy scale of the QSL is very small, due to
tunneling only appearing at high order in perturbation theory,
confining its effects very low temperatures and close proximity to the
\SI{} point \cite{banerjee-isakov-2008-unusual, kato-onoda-2015}.  In
the \SS{}, quantum dynamics appear at \emph{first order} in
perturbation theory \cite{supp}, and thus we expect them to be more
significant than in \SI{}. The presence of these first order matrix
elements is directly reflected in the presence of single-spin-flip and
spin-exchange dynamics of the \SS{} manifold. Even with such mixing,
when the perturbed Hamiltonian is projected into the \SS{} manifold it
still breaks up into infinitely many disconnected blocks, representing
sets of states reachable by such local moves. The simplest blocks
correspond to a small number of dynamical clusters well-separated by
frozen regions. For example, there can be many $n_c=1$ clusters as in
Fig. \ref{fig:dynamics}, each with two states, corresponding to the
freely flippable spin $\ket{\up}$ and $\ket{\down}$ for each
cluster. Application of a transverse field $\sim -\Gamma \sum_i
\sigma^x_i$ mixes the two states and gives a ground state of
$(\ket{\up}+\ket{\down})/\sqrt{2}$ with energy gain of $-\Gamma$ per
dynamical spin.  Other blocks correspond to more complicated dynamical
clusters with more spatially extended structures.  For example, for
the large linear clusters discussed in the \suppinfo{} \cite{supp} the
energy gain per dynamical spin is smaller, approaching $\sim
-2\Gamma/n_c$ for clusters of size $n_c$ \cite{supp}.  More
exotically, one can even construct states where a single dynamical
cluster of size $n_c \sim O(N)$ encompasses nearly all of the spins in
the system.  Similar considerations apply for transverse exchange
$-J_{\perp} \sum_{\avg{ij}}\left(\s^+_i \s^-_j + \s^-_i
\s^+_j\right)$.  A key difference is that odd-sized dynamical clusters
are guaranteed to have degenerate ground states due to Kramers'
theorem. In the exchange case, the $n_c=1$ clusters thus remain free
spins and gain no energy.

We thus conclude that for quantum SS, the ground states favoured at
first-order in perturbation theory will depend on the ground state
energies of this zoo of clusters as well as their effective packing
fractions.  We leave the detailed resolution of these non-trivial
questions to future work. As this model is free of the sign problem,
some of these questions should be addressable through quantum Monte
Carlo simulations for both a ferromagnetic transverse exchange
($J_{\pm} >0$) or an arbitrary transverse field.  The physics of the
above dynamical clusters and the heterogeneous freezing could
potentially enlighten our understanding of the phenomena of persistent
spin dynamics \cite{mcclarty-gingras-cosman-2011}. In a more concrete
setting, one may speculate that the \SS{} could be connected to the
physics observed in the QSL candidate \abo{Tb}{Ti}.  A tantalizing
clue are the short-range correlations
\cite{fritsch-ross-2013-spin-ice} at wave-vector \hhh{} seen in
\abo{Tb}{Ti} and the \hhh{} phase obtained by perturbing \SS{}.

In summary, we have identified ``spin slush'', a new cooperative
paramagnet on the pyrochlore lattice found by extending spin ice with
further neighbour exchanges.  This classical Ising model serves as a
simple example of freezing and dynamical heterogeneity in a clean,
disorder-free system. The features present in the magnetic
correlations and the unusual low temperature dynamics could prove
useful in understanding such physics in real materials.

\acknowledgements{ We thank Yuan Wan for helpful comments and
  discussions.  This work was supported by the NSERC of Canada, the
  Canada Research Chair program (M.G., Tier 1), the Canadian
  Foundation for Advanced Research and the Perimeter Institute (PI)
  for Theoretical Physics. Research at PI is supported by the
  Government of Canada through Industry Canada and by the Province of
  Ontario through the Ministry of Economic Development \& Innovation.
}

\appendix
\section{Details of Monte Carlo simulations}
For all Monte Carlo simulations we used the standard Metropolis
updating scheme with single spin flip moves. For thermodynamic
quantities we simulated systems of $N=16L^3$ spins in $L^3$
conventional cubic unit cells of the pyrochlore lattice under periodic
boundary conditions with linear size up to $L=10$. Typically, we used $O(10^6)$
sweeps to anneal the system to each temperature and thermalize, then
an additional $O(10^6)$ sweeps were used to compute observables. Error
estimates were computed via the bootstrap method. For spin-spin and
charge-charge correlation functions, we simulated larger systems of
size up to $L=24$, but only $O(10^5)$ sweeps were needed to obtain 
accurate results. In both cases, we also used parallel tempering moves
after each sweep to aid equilibriation. Longer simulations on smaller
system sizes, with $O(10^7)$ to $O(10^8)$ sweeps produce results
consistent with the shorter simulations on the larger systems. For
dynamical quantities, a comparable number of sweeps and system sizes
were used, except without the use of parallel tempering. To access
the very low temperature auto-correlation function, we first slowly
annealed the system to $T/J=10^{-4}$, guaranteeing that an \SS{}
ground state was reached, then followed the same protocol as the
higher temperature simulations. This was repeated many times; two of
these simulations are described in the main text.

\bibliography{draft}

\clearpage

\addtolength{\oddsidemargin}{-0.75in}
\addtolength{\evensidemargin}{-0.75in}
\addtolength{\topmargin}{-0.725in}

\newcommand{\addpage}[1] {
    \begin{figure*}
      \includegraphics[width=8.5in,page=#1]{supp.pdf}
    \end{figure*}
}

\addpage{1}
\addpage{2}
\addpage{3}
\addpage{4}
\addpage{5}
\addpage{6}
\addpage{7}

\end{document}